\documentclass[twocolumn,printnumbers,amsmath,amssymb,showpacs,10pt]{revtex4-1}

\newcommand{\be}{\begin{equation}}
\newcommand{\ee}{\end{equation}}
\newcommand{\ba}{\begin{eqnarray}}
\newcommand{\ea}{\end{eqnarray}}
\usepackage{graphicx}
\usepackage{dcolumn}
\usepackage{bm}
\usepackage{ulem}
\usepackage{color}
\begin{document}

\title{{Transmission} and reflection of strongly nonlinear solitary waves at granular interfaces}
\author{A. M. Tichler$^{\dag,*}$, L. R. Gomez $^{\dag,**}$, N. Upadhyaya $^{\dag}$,  X. Campman$^{*}$, V. F. Nesterenko$^{\dag\dag}$ and V. Vitelli$^{\dag}$}
\affiliation{$^{\dag}$ Instituut-Lorentz for Theoretical Physics, Universiteit Leiden, 2300 RA Leiden, The Netherlands \\
$^{*}$Shell Global Solutions International B.V., Kessler Park 1, 2288 GS, The Netherlands \\
$^{**}$ Department of Physics, Universidad Nacional del Sur - IFISUR - CONICET, 8000 Bah\'ia Blanca, Argentina\\
$^{\dag\dag}$ Jacobs School of Engineering, University of California San Diego, 9500 Gilman Drive, La Jolla, USA.}

\begin{abstract}
\noindent

The interaction of a solitary wave front with an interface formed by two strongly-nonlinear non-cohesive granular lattices displays rich behaviour, characterized by the breakdown of continuum equations of motion in the vicinity of the interface. By treating the solitary wave as a quasiparticle with an effective mass, we construct an intuitive (energy and linear momentum conserving) discrete model to predict the amplitudes of the transmitted solitary waves generated when an incident solitary wave front, parallel to the interface, moves from a denser to a lighter granular hexagonal lattice. Our findings are corroborated with simulations.  We then successfully extend this model to oblique interfaces, where we find that the angle of refraction and reflection of a solitary wave follows, below a critical value, an analogue of Snell's law in which the solitary wave speed replaces the speed of sound, which is zero in the sonic vacuum.  
\\
\\

\end{abstract}

\maketitle

The study of solitary waves has over the years led to a
paradigmatic shift in our understanding of many body phenomena
characterized by anharmonic effects that manifest themselves in
exotic electronic \cite{Heeger} and mechanical states
\cite{Nesterenko_Book,Sen0}. A concrete and technologically
relevant \cite{Daraio,DaraioII} arena to study strongly non-linear
mechanical waves is the {\it sonic vacuum}
\cite{Nesterenko_1995,Nesterenko_1994} -- a paradigmatic example of
which is an aggregrate of grains just in contact. Owing to
the vanishing speed of linear sound, even the tiniest strains
propagate as supersonic solitary waves, non-linear periodic waves
and shock-like waves depending on conditions of loading. However, the differential equations
describing the propagation of mechanical disturbances around the state of sonic vacuum are both
non-linear and generally not integrable, making it difficult to model solve them analytically. Moreover, the continuum
approximation itself may fail in the vicinity of a sharp granular
interface, where the discrete nature of the granular medium
dominates.

So far most studies of strongly  non-linear granular interfaces have concentrated on what happens when a solitary wave initially propagating in a chain with mass $m_1$ reaches an interface where the particles mass suddenly changes to $m_2$. Depending on the ratio $A=\frac{m_2}{m_1}$, qualitatively different behaviours are observed. 
When the solitary wave moves from a lighter to a denser medium ($A>1$), most of its energy gets divided into a reflected and a transmitted pulse, whose respective amplitudes can be estimated using the conservation of linear momentum and energy \cite{Nesterenko_1995,Nesterenko_1994,Nesterenko_Book}. By contrast, when the incident solitary wave moves from a denser to a lighter medium ($A\ll1$), a train of (multiple) solitary waves is generated in the lighter medium \cite{Nesterenko_Book,Nesterenko_1995}. In this case, a direct application of the two conservation laws (momentum and energy) is not sufficient to predict the amplitude ratios of the {\it multiple} solitary waves.
Experimental studies of granular chains comprised of steel and polytetrafluoroethylene (PTFE) particles ($A=0.27$) have shed light on the discrete mechanism responsible for the generation of the train of solitary waves in the PTFE light chain. Most of the collective motion carried by the incident solitary wave (propagating in the chain of stainless steel particles) is converted into the motion of a single interfacial steel particle \cite{Nesterenko_2005,Job,Sen_2007}. 

In this Letter, we turn to the hitherto unexplored two dimensional problem of determining the reflection and transmission of a strongly non-linear solitary wave-front incident upon an interface between two hexagonal lattices both in a sonic vacuum, but with different particle masses. We treat the solitary waves as quasiparticles with an effective mass and model the interaction with a two-dimensional granular interface, by assuming an energy and linear momentum conserving scenario validated by simulations. In the $A\ll1$ case, the last row of ``heavy" interfacial beads absorbs on a ``fast" time scale the main part of the energy and linear momentum of the incident solitary wave-front (assumed parallel to the interface) and subsequently decelerates on a ``slow" time scale, generating a train of (asymptotically) separated solitary waves in the ``lighter" sonic vacuum. Crucial to understanding this phenomenon is the role of contact breaking at the interface and the resulting break-down of the continuum approximation. When a strongly non-linear wave is incident at an oblique angle to the interface, we find that the angles of refraction and reflection are surprisingly well captured by a granular analogue of Snell's law that holds irrespective of the solitary wave-front amplitude.

\begin{figure}
\begin{center}
\includegraphics[width=0.5\textwidth]{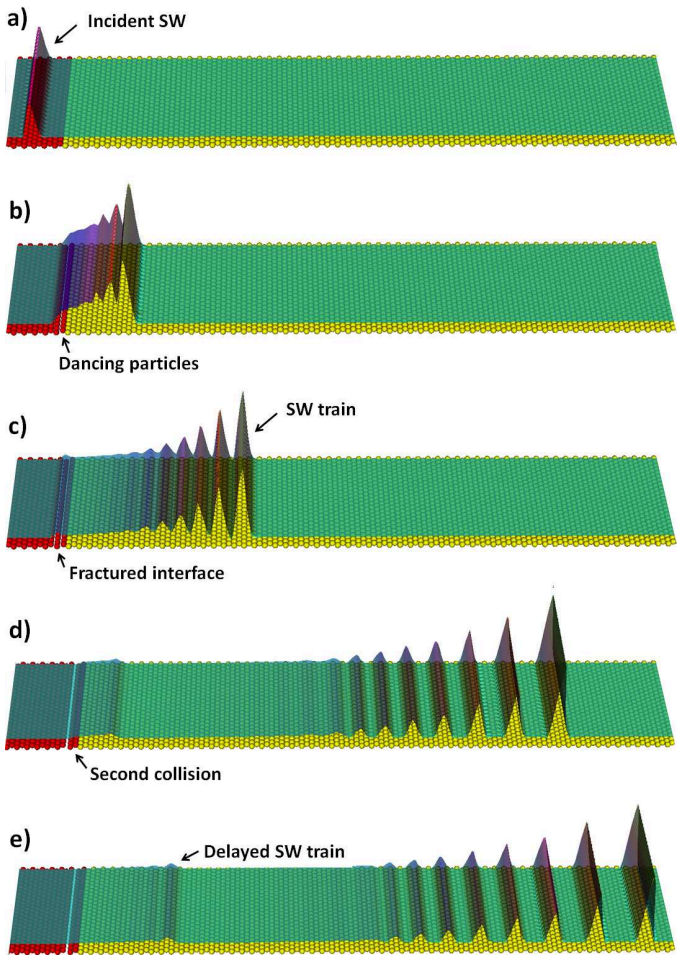}
\caption{Time sequence leading to the generation of a solitary wave train in simulations. The (red) beads on the left of the interface constitute the heavier medium with mass $m_1$ and the (yellow) beads on the right of the interface constitute the lighter medium with mass $m_2$. The mass ratio $A\equiv \frac{m_2}{m_1}=0.125$. The velocity field overlayed in green, denotes the instantaneous speeds of the beads.}
\label{SWT}
\end{center}
\end{figure}

\textit {Simulations of parallel interfaces.}  In order to investigate solitary waves scattering at two dimensional granular interfaces, we performed
molecular dynamics simulations of an impact experiment performed on hexagonal packings of $10^4$
frictionless spherical grains. As shown in Fig. \ref{SWT}, an interface is introduced by assigning a
mass $m_1$ to the rows of grains on its left (shown in red) and a mass $m_2$ to rows on its right (shown in yellow). Both portions of the hexagonal lattice are comprised of grains with zero initial overlap and equal diameters.
Two grains of radius $R$ and masses $\{m_i,m_j\}$ at positions $\{\vec{x}_i,\vec{x}_j\}$ interact via a one-sided non-linear repulsive potential following Hertz law \cite{Landau_Book}
\begin{equation}
V_{ij}=\frac{K_{ij}}{\alpha}\,\,\delta_{ij}\,^{\frac{5}{2}}
\label{eq:potential}
\end{equation}
only for positive compressional strains $\delta_{ij}\equiv  2 R -|\vec{x}_i
-\vec{x}_j|>0$, otherwise  $V_{ij}=0$, when $\delta_{ij}\le0$. Here, the
interaction parameter $K_{ij}=\frac{2}{3} R E^{*}_{ij}$ is expressed in terms of the effective
Young's modulus of the two particles, $E^{*}_{ij}$, see Ref.
\cite{Somfai} for more details.
At $t=0$ we impart to the left-most row a speed $u_{p}$ and subsequently integrate Newton's equations of motion
numerically subject to periodic boundary conditions perpendicularly to the direction of propagation.

As shown in Fig. \ref{SWT}(a), this initial condition leads to the generation of a non-linear wave front parallel to the interface traveling towards the right with
a speed $V_s \sim u_p^{1/5}$ analogously to solitary waves in granular chains \cite{Nesterenko_Book}.   
At later times shown in Fig. \ref{SWT}(b), when the solitary
wave has interacted with the interface, we see a ruptured
interface with one of the interfacial rows of heavy (red) beads ``dancing" in
contact with the lattice of lighter (yellow) beads, throttling the
generation of an oscillatory wave profile in the lighter lattice
close to the interface.  This oscillatory wave is subsequently disintegrated
into a sequence of separate solitary waves, as shown in Fig. \ref{SWT}(c). 
The separate solitary
waves
propagate with different speeds (dependent on their amplitude), while a second collision of the 
``dancing" interfacial row of particles, shown in Fig. \ref{SWT}(d), generates 
a second delayed solitary-wave train with smaller amplitudes, see Fig. \ref{SWT}(e) and movie 1 in SI. 

\textit {Quasiparticle Collision Model.} We take advantage of the isotropic elasticity of the hexagonal lattice to assume that the dynamics of a solitary wave-front parallel to the interface, as in Fig. \ref{SWT}, is effectively one dimensional and governed, in the continuum limit, by the non-linear wave equation \cite{Nesterenko_Book}
\begin{equation}
\xi_{tt}=c^2 \left[\xi^{\frac{3}{2}}+\frac{2R^2}{5}\xi^{\frac{1}{4}}(\xi^{\frac{5}{4}})_{xx}\right]_{xx} , \label{fg}
\end{equation}
where $c$ is a material constant and $\xi(x,t)$ is  the strain field
$\xi(x,t)=-\partial_x u(x,t)$ expressed in terms of the particle displacement field, $u(x,t)$, along the x direction. The left-hand side of Eq. (\ref{fg}) is the standard inertia term, the second term on the right-hand side captures non-linear dispersive effects, while the first arises from the restoring force as in the wave-equation, if one considers that the force is not linear, but it depends on $\xi^{3/2}$ according to Hertz law. 

A  strongly nonlinear solitary wave solution of Eq. (\ref{fg}) can be derived analytically \cite{Nesterenko_Book} and it has been validated by extensive simulations and experiments mostly on granular chains \cite{Nesterenko_1994,Nesterenko_Book,Coste2,Daraio2005,Job,Sen,Gomez_2012,Gomez2}.  Crucially, the total energy $E=\frac{P^2}{2m_{\text{eff}}}$ carried by the solitary wave depends quadratically on its total momentum $P$, which allows to define an effective mass $m_{\text{eff}} \approx 1.4 m$ for the solitary wave \cite{Nesterenko_1995,Nesterenko_1994,Nesterenko_Book,Job,Hinch,Jin}.

\begin{figure}
\begin{center}
\includegraphics[width=0.45\textwidth]{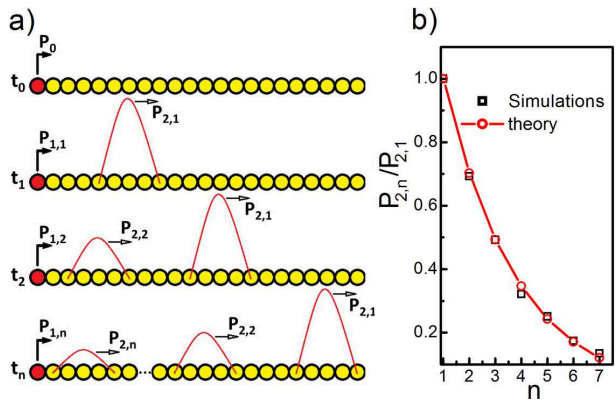}
\caption{(a) Schematic illustration of our model for the formation of a solitary-wave train, side-view. (b) Momentum ratios $\frac{P_{2,n}}{P_{2,1}}$  between the $n$-th solitary wave and the leading one in the train for $A=\frac{m_2}{m_1}=0.125$. Red circles are the theoretical predictions while the black squares are the numerical values from the simulations of Fig. 1.} \label{model}
\end{center}
\end{figure}

The notion of the solitary wave as a quasiparticle allows to construct a simple quasi one-dimensional model for the generation of the solitary wave train, illustrated schematically in Fig. \ref{model}. At $t_0$, we assume that a chain of {\it light}
yellow beads is uncompressed and all the energy and linear momentum $P_0$, carried by the incident
solitary wave, is concentrated in the {\it heavy} red interfacial
particle. At a subsequent time $t_1$ a
single solitary wave is generated in the {\it light} chain by
reducing the energy and linear momentum of the interfacial heavy
particle. We apply conservation of energy and linear momentum to
the collision process between the ``dancing" bead with mass $m_1$ and the solitary wave, treated as
a quasiparticle with mass $m_{2,\text{eff}}$. We calculate the momentum of the ``dancing"
interfacial particle $P_{1,1}$ after the first collision as 
\begin{eqnarray}
P_{1,1}= \frac{P_0\left(B-1\right)}{\left(B+1\right)}  ,
\label{momentum1}
\end{eqnarray}
where $B \equiv
\frac{m_1}{m_{2,\text{eff}}}$. The momentum $P_{2,1}$ carried by the first leading solitary wave at $t=t_1$ is
\begin{eqnarray}
P_{2,1}=\frac{2P_0}{B+1} . \label{momentum2}
\end{eqnarray}
At time $t_2$
another independent single solitary wave is generated in the
"light" chain, further reducing the energy and linear momentum
of the "dancing" interfacial particle. Upon applying conservation of
energy and linear momentum, as before, and assuming that the first solitary wave does not participate in
this process, we find the momentum of the ``dancing" particle at $t=t_2$, $P_{1,2}$, and of the second
solitary wave, $P_{2,2}$, as
\begin{eqnarray}
P_{1,2}= \frac{P_0\left(B-1\right)^2}{\left(B+1\right)^2},
P_{2,2}=\frac{2P_0 \left(B-1\right)}{\left(B+1\right)^2}. \label{momentum}
\end{eqnarray}
Upon iterating this process $n$ times 
, we
find that the "heavy" interfacial bead at $t=t_n$ is left with a linear
momentum $P_{1,n}$ while the $n$-th solitary wave carries momentum $P_{2,n}$ 
given by
\begin{eqnarray}
P_{1,n}= \frac{P_0\left(B-1\right)^n}{\left(B+1\right)^n},
P_{2,n}=\frac{2P_0\left(B-1\right)^{n-1}}{\left(B+1\right)^n}.
\end{eqnarray}

Fig \ref{model}(b) illustrates the favorable comparison of
$\frac{P_{2,n}}{P_{2,1}}= \left(\frac{B-1}{B+1}\right)^{n-1}$ against numerical data (red circles) for
$A=0.125$. The amplitudes of the delayed secondary
sequence of solitary waves generated is neglected in our approximate model.

\begin{figure}
\vspace{0pt}
\includegraphics[width=0.47\textwidth]{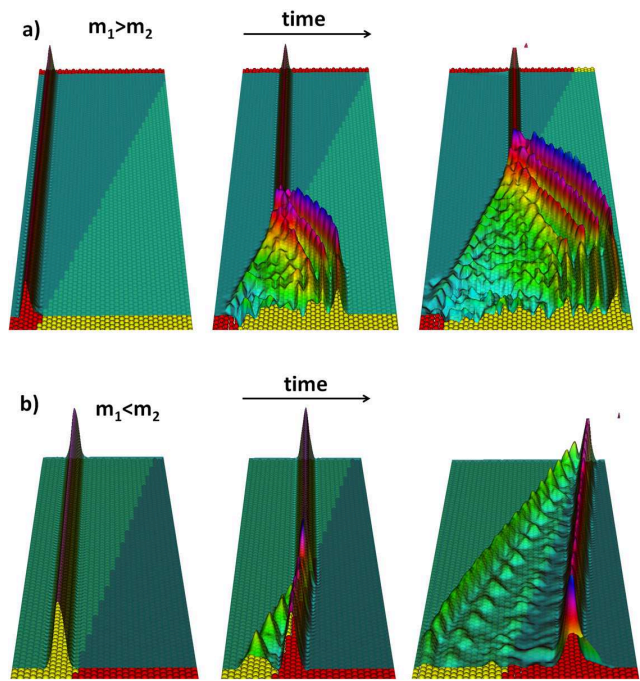}
\caption{Snapshot of the simulations showing a solitary wave incident upon an interface separating two hexagonal lattices in a sonic vacuum. (a) For $A\equiv \frac{m_2}{m_1}=0.125$, the transmitted disturbance propagates in the form of a non-linear oscillatory wave analogous to the train of solitary waves shown for granular chains in Fig. \ref{SWT} for $A<1$. (b) For $A>1$, we find both a reflected and transmitted solitary wave as shown in figure for $A=3$. 
}
\label{2D}
\end{figure}

\begin{figure}
\vspace{0pt}
\includegraphics[width=0.5\textwidth]{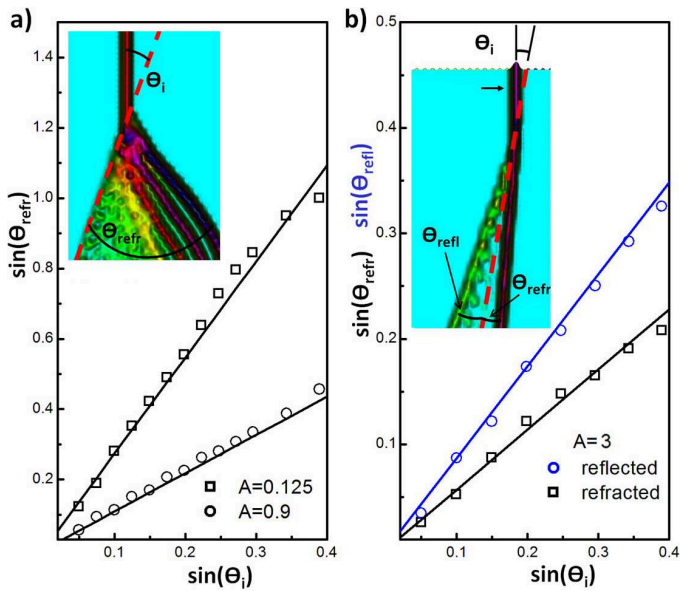}
\caption{(a) Angle of refraction $\theta_{\text{refr}}$ vs angle of incidence $\theta_i$ for the hexagonal lattice when $A<1$. The square (circle) symbols correspond to the case when $A=0.125$($A=0.9$) and compares  the numerically obtained ratio of the speed of the leading transmitted solitary wave to the incident solitary wave, against the analytical estimates given by solid curves. (b) Comparison of numerical data (symbols) with the analytical estimate (solid curves) for the angle of refraction ( black data )  and angle of reflection (blue data) vs the angle of incidence  for the hexagonal lattice when $A>1$. The insets to (a) and (b), describes the relevant angles, where the interface is shown as the dashed (red) line and arrows represent the direction of propagation of the solitary wave front (thick dark region).}
\label{plot}
\end{figure}

\textit{Oblique interfaces}-- The simulation snapshots in Fig. \ref{2D} illustrate the propagation and interaction of a solitary wave-front with an oblique interface separating two hexagonal granular lattices for the case $A\equiv \frac{m_2}{m_1}<1$ in panel (a) and for the case $A>1$ in panel (b) -- see also movies 2a and 2b in SI. 
From these simulations, we have determined numerically the angle of refraction $\theta_{\text{refr}}$ for different values of the angle of incidence $\theta_{i}$, as shown in the inset of Fig. \ref{plot}(a). 

Inspection of the main panel of Fig. \ref{plot}(a), suggests that a linear relationship exists between $\sin{\theta_{\text{refr}}}$ and $\sin{\theta_{\text{i}}}$ for mass ratios $A\ll1$ (squares) and $A\le1$ (circles). We now show that the measured proportionality coefficient is consistent with a non-linear analogue of Snell's law $V_0 \sin{\theta _{\text{refr}}}= V_{2,1}\sin{\theta _{i}}$ where $\{V_{2,1}, V_0\}$ denote respectively the speeds of the (leading) refracted solitary wave and of the incident one. 
To work out explicitly the dependence of $\frac{V_{2,1}}{V_0}$ on the mass ratio $A<1$, we employ a reasoning similar to the one leading to Eq.\ (\ref{momentum2}) that accounts for the discrete mechanism at play near the interface. Upon making use of the scaling relation $P \sim m^3 {V_s}^5$ between the total momentum $P$ carried by a solitary wave and its speed $V_s$ \cite{Job}, we find in the limit of small $\theta_i$ that

\begin{equation}
\frac{\sin(\theta_{\text{refr}})}{\sin(\theta_i)} \approx
\left(\frac{2}{(B+1)A^3}\right)^{\frac{1}{5}}.
\label{eqrefr}
\end{equation}

The right hand-side of Eq.\ (\ref{eqrefr}) is the slope of the continuous line plotted for $A=0.125$ and $A=0.9$ in Fig. \ref{plot}(a). It matches the data (open symbols) obtained by numerically evaluating the left hand-side of Eq.\ (\ref{eqrefr}). This agreement corroborates the non-linear analogue of Snell's law. Note, for the case $A<1$, the reflected wave amplitude is negligibly small, see Fig.\ \ref{2D}(a).

By contrast in the case  $A>1$, Figure \ \ref{2D} (b) shows a train of reflected waves with most of the energy concentrated in the leading solitary wave, which allows us to neglect rupturing at the interface.  As shown in Fig.\ \ref{plot} (b), the angle of reflection $\theta_{\text{refl}}$ is not equal to the angle of incidence $\theta_i$, since the reflected solitary wave speed $V_{\text{refl}}$ is not the same as the incident speed $V_0$. Instead, we evaluate the  ratio $\frac{V_{1,1}}{V_0}$ in analogy with the derivation of Eq. (\ref{eqrefr}). Since in this case, the appropriate collision model is that between two solitary wave quasiparticles, the effective mass contribution cancels out and we obtain an equation analogous to Eq.\ (\ref{momentum1}) with the replacement $B\rightarrow 1/A$ and a sign reversal. Thus,   Snell's law for reflection assumes the form 
\begin{equation}
\frac{\sin(\theta_{\text{refl}})}{\sin(\theta_i)} \approx \left(\frac{A-1}{A+1}\right)^{\frac{1}{5}}.
\label{refle}
\end{equation}
The right hand-side of  Eq.\ (\ref{refle}) is the slope of the continuous (blue) line plotted for $A=3$ in Fig. \ref{plot}(b) and matches the numerical data (blue circles) obtained from evaluating the left hand-side of Eq.\ (\ref{refle}). A similar agreement is found also for the angle of refraction in this case, as illustrated by the data (black squares) and (black) continuous line in Fig. \ref{plot}(b). Despite the fact that our granular lattices are in a state of sonic vacuum, the ratio between the sine of the angles in Snell's law is nearly independent of the amplitude of the incident solitary wave front (see Figure S1 in SI). This seemingly puzzling observation can be rationalized by viewing the incident and (leading) reflected or refracted solitary waves as quasiparticles scattering at the interface whose speeds are proportional
to the incoming ones.

Snell's law also implies the existence of a critical angle of incidence, $\theta_{c}$, for which the transmitted solitary wave will propagate in a direction parallel to the line of the interface or $\theta_{\text{refr}}=90^\circ$ in Eq.\ (\ref{eqrefr}), with the replacement $B\rightarrow1/A$.
In this case, the fronts of the incoming and transmitted solitary wave are no longer continuous over the interface. Moreover, the incoming and reflected solitary waves cross each other at the interface, unlike the case when $\theta_{\text{refr}}<90^\circ$, when the two just touch each other. Figure S2 in SI shows the delayed reflection phenomenon that occurs when the angle of incidence is greater than $\theta_{c}$.

\textit{Conclusion.} We have constructed a discrete model that predicts the amplitudes of transmitted and reflected solitary wave-fronts from a 2D granular interface. We find that the angle of refraction and reflection follow an analogue of Snell's law in which the solitary wave speed replaces the vanishing speed of sound.
\\
\\
\textbf{Acknowledgments} We acknowledge discussion with S. Ulrich. LRG and NU acknowledge financial
support from FOM, Shell and CONICET. AMT, LRG and NU contributed equally to this work.

\end{document}